\documentclass[aps,reprint,floatfix,superscriptaddress]{revtex4-1}

\usepackage[usenames,dvipsnames]{xcolor}
\usepackage{graphicx}
\usepackage{tikz}
\usepackage{datetime}

\begin{document}
\title{Electronic structure of the negatively-charged silicon-vacancy center in
diamond}

\author{Lachlan J. Rogers}
\email{lachlan.rogers@uni-ulm.de}
\author{Kay D. Jahnke}
\affiliation{Institut f\"ur Quantenoptik and IQST, Universit\"at Ulm, D-89081
Ulm, Germany}
\author{Marcus W. Doherty}
\affiliation{Laser Physics Centre, Research School of Physics and Engineering,
Australian National University, ACT 0200, Australia.}
\author{Andreas Dietrich}
\author{Liam McGuinness}
\author{Christoph M\"uller}
\affiliation{Institut f\"ur Quantenoptik and IQST, Universit\"at Ulm, D-89081
Ulm, Germany}
\author{Tokuyuki Teraji}
\affiliation{National Institute for Materials Science, 1-1 Namiki, Tsukuba,
Ibaraki 305-0044, Japan}
\author{Hitoshi Sumiya}
\affiliation{Advanced Materials R\,\&\,D Laboratories, Sumitomo Electric Industries Ltd, Itami, Hyogo 664-0016, Japan}
\author{Junichi Isoya}
\affiliation{Research Center for Knowledge Communities, University of Tsukuba,
1-2 Kasuga, Tsukuba, Ibaraki 305-8550, Japan}
\author{Neil B. Manson}
\affiliation{Laser Physics Centre, Research School of Physics and Engineering,
Australian National University, ACT 0200, Australia.}
\author{Fedor Jelezko}
\affiliation{Institut f\"ur Quantenoptik and IQST, Universit\"at Ulm, D-89081
Ulm, Germany}

\newcommand{\todoANU}[1]{ {\color{Magenta}\textsc{#1}} }
\newcommand{\todoULM}[1]{ {\color{ForestGreen}\textsc{#1}} }

\newcommand*\circled[1]{\tikz[baseline=(char.base)]{
  \node[shape=circle,draw,inner sep=1pt] (char) {#1};}}

\begin{abstract}
The negatively-charged silicon-vacancy (SiV$^-$) center in diamond is a
promising single photon source for quantum communications and information
processing.  However, the center's implementation in such quantum technologies
is hindered by contention surrounding its fundamental properties.  Here we
present optical polarization measurements of single centers in bulk diamond
that resolve this state of contention and establish that the center has a
$\langle111\rangle$ aligned split-vacancy structure with $D_{3d}$ symmetry.
Furthermore, we identify an additional electronic level and evidence for the
presence of dynamic Jahn-Teller effects in the center's 738 nm optical
resonance.
\end{abstract}
\pacs{61.72.jn, 71.55.Cn, 81.05.ug}


\keywords{silicon vacancy, diamond, spectroscopy, polarization, electronic structure}

\maketitle

\section{Introduction}

Single quantum emitters in solids are promising sources of single photons
\cite{kurtsiefer2000stable, brouri2000photon}, architectures for qubits
\cite{childress2006fault-tolerant}, and biological probes
\cite{fu2007characterization, Vaijayanthimala2009functionalized,
ermakova2013detection}.  The negatively-charged nitrogen-vacancy (NV$^-$) color
center in diamond is a prominent example \cite{doherty2013nitrogen-vacancy},
but its optical fluorescence is spread over a large phonon sideband with only
4\% present in the zero phonon line (ZPL).  In contrast, the negatively-charged
silicon-vacancy (SiV$^-$) center in diamond is known to have a small phonon
sideband, with 70\% of its fluorescence concentrated in a ZPL at 738\,nm (1.68
eV) \cite{collins1994annealing}.  This is a marked advantage for applications
that require indistinguishable photons, such as quantum communication and
information processing architectures that rely on photon entanglement
\cite{benjamin2009prospects}.  It is also beneficial for technological
applications involving cavity QED \cite{neu2013low-temperature, wang2006single,
neu2011single}.  However, persistent contention surrounding the fundamental
properties of the SiV$^-$ center has hindered its implementation in such
quantum technologies.

This contention arises mainly from conflicting reports of the geometrical
alignment of SiV$^{-}$ within the diamond lattice.  Optical polarization
measurements have suggested the center is aligned along $\langle110\rangle$
crystal vectors \cite{brown1995site,neu2011fluorescence}, while a
$\langle111\rangle$ alignment was concluded from \textit{ab initio}
calculations \cite{goss1996twelve-line} and related electron paramagnetic
resonance (EPR) measurements of the neutral charge state (SiV$^0$)
\cite{dhaenens-johansson2011optical}. Further disagreement is
added by a past observation of the Zeeman splitting of the 738 nm ZPL fine
structure, which suggests that the center has a $\langle100\rangle$
orientation \cite{sternschulte1995uniaxial}.

Here we report a study of the polarization of photoluminescent emission (PL)
and excitation (PLE) of single SiV$^-$ centers in bulk diamond at room and
cryogenic temperatures, and confirm the $\langle111\rangle$ alignment.  Our
study avoids the limitations of the previous polarization studies by utilizing
single center interrogation in bulk samples which have well defined
crystallographic orientation and low intrinsic strain.  This alignment is
consistent with a silicon split-vacancy structure having $D_{3d}$ symmetry, and
our conclusions are supported by recent Zeemann studies
\cite{hepp2014electronic}.  Additionally, by examining the polarization
properties of the PL fine structure we have confirmed that the 738\,nm ZPL
arises from a $^2\!E_g$$\leftrightarrow$$^2\!E_u$ transition and that the fine
structure arises from spin-orbit interactions. Analysis of the PL polarization
of the accompanying phonon sideband strongly suggests dynamic Jahn-Teller
effects that invite future investigation. Furthermore, our PLE polarization
observations reveal an additional $^2\!A$ electronic level, which is excited at
wavelengths $\lesssim605$\,nm ($\gtrsim2.05$\,eV). Based upon past PLE spectra
\cite{iakoubovskii2000luminescence} and the electronic model of the center
\cite{goss1996twelve-line,dhaenens-johansson2011optical}, this level is
precisely identified as $^2\!A_{1g}$.  These observations provide the most
complete understanding of the SiV$^-$ electronic structure yet obtained.


\section{Polarization of photoluminescence}

Two samples with SiV$^-$ densities of $\sim0.1\,\mu\mbox{m}^{-3}$ were used in
our experiments: one with a $\{111\}$ face and one with a $\{100\}$ face.  The
first sample was a low strain high-pressure high-temperature (HPHT) diamond
that was laser cut to provide the desired $\{111\}$ surface.  The SiV$^-$ sites
found in this surface were naturally occuring and must have been formed during
the HPHT growth process.  The $\{100\}$ sample was a microwave plasma-assisted
chemical-vapor-deposition (MPCVD) film grown on a $\langle100\rangle$-oriented
plate cut from a low-strain, type-IIa, HPHT crystal (Sumitomo Electric
Industries, Ltd.).  Details of the MPCVD apparatus have been described
elsewhere \cite{teraji2012chemical}.  Si atoms were
introduced into the growth plasma as it etched a 6H-SiC single-crystal plate which was
placed in the sample chamber. SiC has a greater resistance to hydrogen plasma
than both Si and SiO$_2$, which slowed the Si incorporation rate enough to
produce SiV$^-$ centers with a density that allowed individual sites to be
identified. 

\begin{figure}[thb]
\includegraphics[width=8.6cm]{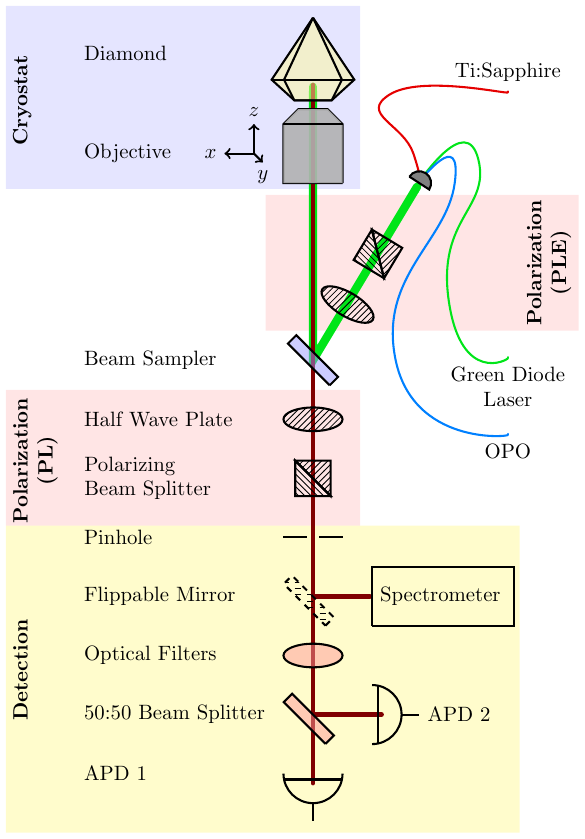}
\caption{
	Experimental setup.  The custom-built confocal microscope was built around
	a continuous flow helium cryostat that held the sample.  A 532\,nm green
	diode laser was used for excitation in the PL measurements, and a
	half-wave-plate and polarizer allowed the fluorescence to be measured as a
	function of polarization angle.  The SiV$^-$ fluorescence (dark red) was then
	measured either using Avalanche Photo Diodes (APDs) or with a Spectrometer.
	PLE measurements were performed using a CW Titanium Sapphire Laser and a
	pulsed Optical Parametric Oscillator (OPO).  In order to measure the
	polarization dependence of the excitation transition, the HWP and polarizer
	were moved from the fluorescence beam to the laser beam at the fibre
	out-coupler. 
}
\label{setup_drawing}
\end{figure}

A green diode laser at 532\,nm was used to excite SiV$^-$ for the
photoluminescence (PL) measurements.  The fluorescence was measured using
either a pair of avalance photo diodes (APDs) or a spectrometer with a
1596\,grooves/mm grating.  The samples were mounted on the cold finger of a
continuous flow helium cryostat capable of cooling them to about 8\,K.  A
custom built confocal microscope (illustrated in Figure \ref{setup_drawing})
was used to study individual color centers, which were confirmed to be single
sites using second-order autocorrelation measurements.  Excitation power was of
order 1\,mW at the microscope objective.  A half wave plate (HWP) and a linear
polarizer were placed in the detection beam so that rotating the HWP allowed
fluorescence to be measured at arbitrary polarization angles.  Having the
linear polarizer fixed and rotating the fluorescence beam with the HWP
eliminated the influence of any polarization dependence of the measurement
aparatus.  

\subsection{$\{111\}$ surface}

\begin{figure}
\includegraphics[trim = 0mm 0mm 0mm 0mm, clip, width=8.6cm]{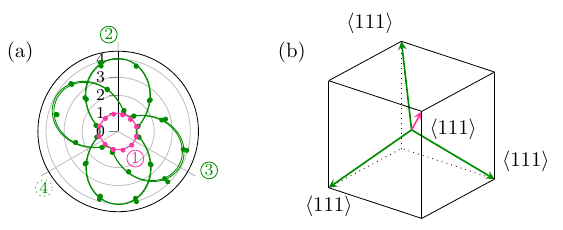}
\caption{
	Emission ZPL intensity as a function of detection polarisation for the
	$\{111\}$ surface.  
	(a) The SiV$^-$ sites formed distinct sets, within which all sites had
	similarly polarized emission. One set had no polarization variation
	($C=3\%$), while two other sets had a contrast of $C=60\%$ and were
	separated by $120^\circ$.
	(b) These sets are arranged in a manner corresponding to the projection of
	the $\langle111\rangle$ crystal vectors, indicating the SiV$^-$ center has
	a $\langle111\rangle$ major symmetry axis.  The position of an expected
	fourth set that should share the contrast of $C=60\%$ is marked, and the
	absence of any SiV$^-$ in this orientation is attributed to crystal growth
	effects.
	}
\label{zpl_polar_plots_111}
\end{figure}

More than 40 fluorescent sites  in the $\{111\}$-faced sample were examined
using spectrometer measurements.  Over half of these sites were found to be
either NV$^-$ or a SiV$^-$ close to a NV$^-$, but 15 pure SiV$^-$ sites were
observed.  Seven typical SiV$^-$ sites lay within a convenient scan region, and
these were measured in further detail.  In particular, the spectrally resolved
ZPL intensity was measured at 8\,K as the detection polarization was rotated.
The resulting ZPL polarizations formed distinct sets as shown in Figure
\ref{zpl_polar_plots_111}(a).

We discuss the polarization dependence in terms of polarization contrast
\begin{eqnarray}
C = \frac{I_{\max}-I_{\min}}{I_{\max}+I_{\min}},
\end{eqnarray}
given as a percentage \cite{clark1971photoluminescence, neu2011fluorescence}.
For the $\{111\}$ surface there was a set of sites (number \circled{1}) which
showed no polarization dependence ($C=3\%$).  Two other sets (numbered
\circled{2} and \circled{3}) each showed a contrast of $C=60\%$ and had their
maxima separated by $120^\circ$.  

These sets correspond to the pattern of $\langle111\rangle$ vectors in the
diamond lattice when viewed through a \{111\} surface, as illustrated in Figure
\ref{zpl_polar_plots_111}(b).  The NV$^-$ sites also formed sets corresponding
to the $\langle111\rangle$ vectors.  However, none of the 40 NV$^-$ and SiV$^-$
sites were observed in orientation \circled{4}, which was the growth direction
of this crystal sector.  There are numerous reports of preferential orientation
when defect sites are incorporated during growth \cite{edmonds2012production,
collins1989polarised, iakoubovskii2004ni-vacancy, iakoubovskii2004alignment,
dhaenens-johansson2010epr, dhaenens-johansson2011optical}.  The absence of
orientation \circled{4} is attributed to similar growth effects, and it does
not impact the strength of our subsequent conclusions.  The exact correlation
of the pattern of SiV$^-$ sites with the pattern of $\langle111\rangle$ crystal
vectors is strong evidence that SiV$^-$ is aligned along these directions in
the lattice.

\subsection{$\{100\}$ surface}

\begin{figure}
\includegraphics[trim = 4.5mm 3mm -3mm 0mm, clip, width=8.6cm]{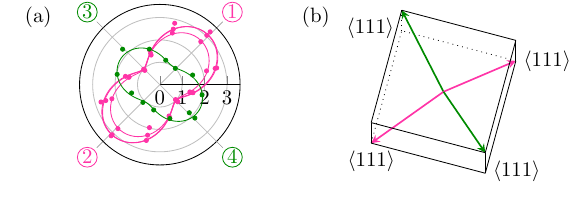}
\caption{
	Emission ZPL intensity as a function of detection polarisation for the
	$\{100\}$ surface.  
	(a) The SiV$^-$ sites formed two distinct sets with contrast of about
	$C=50\%$, separated by $90^\circ$.
	(b) These sets are also arranged in a manner corresponding to the
	projection of the $\langle111\rangle$ crystal vectors.  The difference in
	contrast between the perpendicular sets is attributed to the viewing axis
	being slightly misaligned from $\langle100\rangle$, as is the case in the
	illustrated unit cube.
	}
\label{zpl_polar_plots_100}
\end{figure}

The same measurement was made for 8 SiV$^-$ sites in the $\{100\}$ sample,
resulting in two distinct sets of sites separated by $90^\circ$ as shown in
Figure \ref{zpl_polar_plots_100}(a).  As before, these are consistent with the
pattern of $\langle111\rangle$ vectors in the viewing projection as
illustrated.  Note that since the patterns are perpendicular and the
$\langle111\rangle$ vectors form perpendicular sets, it is also possible to
interpret the patterns as implying that the polarization is orthogonal to
$\langle111\rangle$ (instead of along it).  This ambiguity arises fundamentally
from the geometry, and hinders the $\{100\}$ surface from clearly revealing the
SiV$^-$ dipole properties.  For this reason we focus now on the $\{111\}$
surface for fine-structure measurements.

\subsection{ZPL fine structure}

\begin{figure*}
\includegraphics[width=17.8cm]{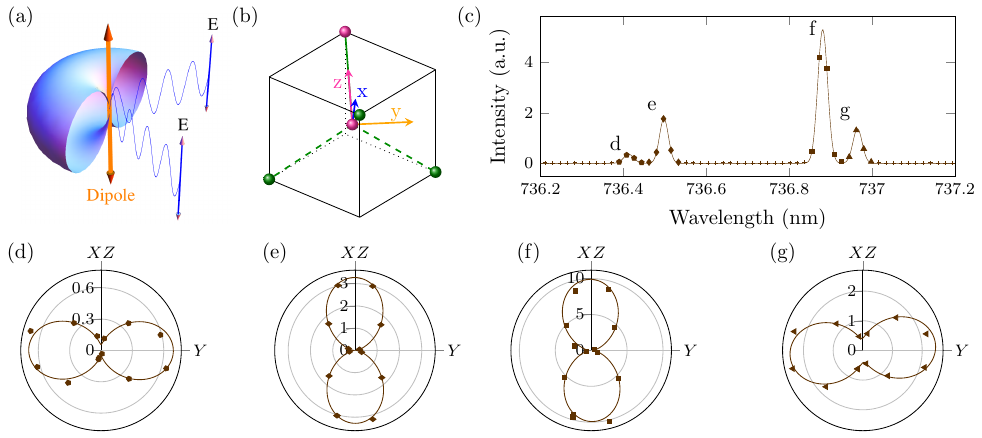}
\caption{
	Polarization of ZPL fine structure.
	(a) A dipole radiates primarily into the plane orthogonal to its axis.  The
	toroidal surface illustrates the relative intensity as a function of
	elevation angle out of this plane, and importantly shows that no radiation
	propogates along the dipole axis. In every direction of propogation, the
	radiation is polarized parallel to the dipole axis (here E indicates the
	electric-field vector and is not a symmetry label). 
	(b) Projection of the local X $[11\overline2]$, Y $[1\overline10]$, and Z
	$[111]$ dipoles for a $\langle111\rangle$ viewing direction.  The X axis
	lies in a symmetry plane, and so each of the three oblique SiV$^-$
	orientations may have Y chosen so that it is perpendicular to the viewing
	direction.  The projection of X is parallel to that of Z. 
	(c) At 8\,K the ZPL is resolved into four components.  The width of these
	lines is limited by the spectrometer resolution.
	(d), (e), (f), (g) The polarization of each ZPL component. The two inner
	lines are polarized in the X\,Z direction, and have almost
	complete contrast ($C_{(e)}=97\%$ and $C_{(f)}=92\%$). The two outer lines
	are oppositely polarized (Y), and have contrasts $C_{(d)}=85\%$ and
	$C_{(g)}=72\%$.
}
\label{zpl_pol}
\end{figure*}

Polarization measurements probe the dipole moments that are active in a given
transition.   In general, an emitter can be fully characterised by considering
three orthogonal dipoles.  The conventional site-referenced axes for a defect
aligned to the $\langle111\rangle$ crystal vectors are Z $[111]$, X
$[11\overline2]$ and Y $[1\overline10]$, and so we will consider dipole moments
along these local axes. A dipole produces light polarized parallel to its axis,
but this light radiates primarily into the plane orthogonal to its axis (Figure
\ref{zpl_pol}(a)).  In the \{111\} surface,  SiV$^-$ orientation \circled{1}
has its Z axis aligned to the viewing direction (Figure
\ref{zpl_polar_plots_111}) .  Since a dipole does not radiate along its axis,
the measured fluorescence can only arise from the X and Y dipole moments.  The
lack of contrast indicates that the X and Y dipole moments have equal
magnitude, denoted $d_\perp$.  The axial dipole moment (along Z) is called
$d_\parallel$.

The other three SiV orientations in the $\{111\}$ sample have their Z axes
inclined at 19.5 degres to the surface.   For these three oblique orientations
there is always a choice of X and Y that makes Y perpendicular to the surface.
The corresponding X then always appears parallel to Z, as illustrated in
Figure~\ref{zpl_pol}(c).  To examine the obliquely oriented sites with clarity
we take X,\,Y,\,Z as conventionally defined, and choose the sample surface to
be $S=(11\overline1)$.  The scalar product allows a quick determination of the
angles that X,\,Y,\,Z form with the surface $S$, making it possible to obtain
the percieved strength of dipole moments along these axes.  Because Y is
perpendicular to the viewing direction the dipole moment in this axis is seen
at 100\%.  Due to the geometry, only 33\% of the dipole moment along X and 94\%
of that along Z are observed.  Thus the expected polarisation contrast for a
transition only involving $d_\perp$ (ie $d_\parallel=0$) is 
$$C_{d_\perp} = \frac{1-0.33}{1+0.33}=50\%$$ 
with maxima in the Y direction.  A transition for which $d_\perp=0$ should have
contrast of $C_{d_\parallel}=100\%$ with maxima in the X\,Z direction (but only
94\% of the intensity will be observed).

At 8\,K the ZPL was spectrally resolved into four components as shown in Figure
\ref{zpl_pol}(a). Photoluminescence from each of the four ZPL components was
measured at various polarizations, and the results are shown against the
projected X\,Z and Y axes in Figures \ref{zpl_pol}(d)-(g).  The inner two lines
are polarized along X\,Z and have a contrast of $C\geq92\%$, while the outer
lines are polarized in the Y direction with contrasts of $C=72\%$ for (d) and
$C=85\%$ for (g).

Since the inner two lines have nearly complete contrast and are polarized in
the X\,Z direction they must arise only from the axial dipole moment
$d_\parallel$.  The outer two lines are polarized in the Y direction, but have
contrast greater than the 50\% expected for a purely $d_\perp$ transition.  For
optical detection normal to the surface, this suggests the Y dipole moment is
stronger than that along X which is in conflict with the observations for
orientation \circled{1}.  However, the microscope objective used here collects
fluorescence within a solid angle around the surface normal.  This amplifies
the apparent strength of the Y dipole moment and leads to higher than expected
contrast.  This effect was assessed by measuring nearby NV$^-$ centers, which
are also $\langle111\rangle$ aligned but have only a perpendicular dipole
moment.  For NV$^-$ sites obliquely angled to the surface, polarization
contrast was found to be $\sim$70\% indicating that SiV$^-$ line (g) in Figure
\ref{zpl_pol} arises purely from $d_\perp$. Line (d) is similar, but as it has
the weakest intensity (and so highest uncertainty in contrast), it is difficult
to draw detailed conclusions.

The magnitudes of the parallel and transverse dipole moments may also be
compared.  It is clear from Figure \ref{zpl_pol} that the inner two lines,
arising from $d_\parallel$, are stronger than the outer lines which arise from
$d_\perp$.  The difference in intensity indicates that $d_\parallel$ is about
four times stronger than $d_\perp$. This indicates that the ZPL is
predominately polarized parallel to $\langle111\rangle$, allowing the SiV
center to be approximated as a single dipole.

\begin{figure}
\includegraphics[width=8.6cm]{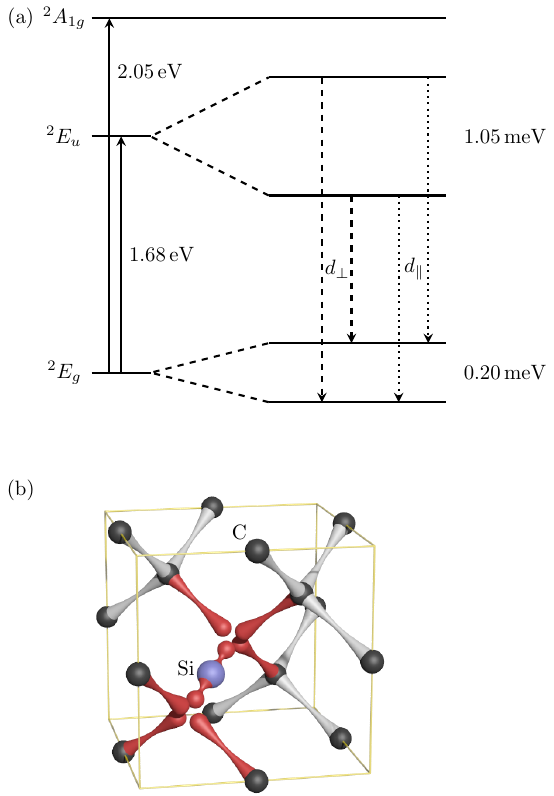}
\caption{
	(a) Electronic structure of SiV$^-$, depicting the ground $^2E_g$ and first
	excited $^2E_u$ levels \cite{goss1996twelve-line}, their doublet fine
	structure \cite{clark1995silicon} and the $^2A_{1g}$ level identified in
	this work. The $^2E_g$$\leftrightarrow$$^2E_u$ transition gives rise to the
	four-line fine structure of the 738\,nm (1.68 eV) ZPL. The depicted
	polarization selection rules are as observed and correspond to a spin-orbit
	origin of the fine structure. Comparison with past PLE spectra
	\cite{iakoubovskii2000luminescence} suggests that the $^2A_{1g}$ level is
	605\,nm (2.05\,eV) above the ground state.
	(b) Schematic of the split-vacancy structure with $D_{3d}$ symmetry.  The
	``dangling bonds'' contributing to the SiV$^-$ centre are highlighted in
	red.  These are drawn schematically to indicate the lattice vacancies, and
	do not correspond to actual electron charge distributions.
}
\label{shape_and_level_scheme}
\end{figure}

Applying symmetry selection rules from group theory, a transition with
differing Z and equal X and Y dipoles at a $\langle111\rangle$ aligned site in
diamond can only occur if the site has $C_{3v}$ or $D_{3d}$ symmetry and if the
transition occurs between two $E$ electronic levels \cite{tinkham1964group}.
This conclusion is consistent with previous interpretations of the four-line
fine structure \cite{sternschulte19941.681-ev, clark1995silicon,
goss1996twelve-line}.  A splitting of the degeneracy in each E state produces
four transitions as shown in Figure \ref{shape_and_level_scheme}. We
observed that the ground and excited state splittings are the same for each of
the single centers we studied and are 0.20\,meV and 1.05\,meV, respectively.
This indicates that these splittings arise from intrinsic properties of
SiV$^-$.  

Based upon our polarization measurements, in Figure
\ref{shape_and_level_scheme}(a) we assign the dipole selection rules of the
fine structure transitions between the $E$ electronic levels.  The observed
selection rules are consistent with a spin-orbit origin of the zero-field
splittings. Within each $^2\!E$ level, the interaction of the electronic $S=1/2$
spin and orbital angular momentum about the $\langle111\rangle$ symmetry axis
leads to a splitting between sub-levels where the angular momenta
constructively and destructively combine.

The fine structure polarization measurements are not able to distinguish
between $C_{3v}$ and $D_{3d}$ site symmetries.  The precise symmetry of the
center may only be established through evidence for the absence ($C_{3v}$) or
presence ($D_{3d}$) of inversion symmetry \cite{tinkham1964group}.  However,
\textit{ab initio} calculations have suggested the $D_{3d}$ symmetry for
SiV$^-$ \cite{goss1996twelve-line}.  This model places the Si atom in the
middle of two adjacent vacancies, as illustrated in Figure
\ref{shape_and_level_scheme}(b).  

\begin{figure}
\includegraphics[width=8.6cm]{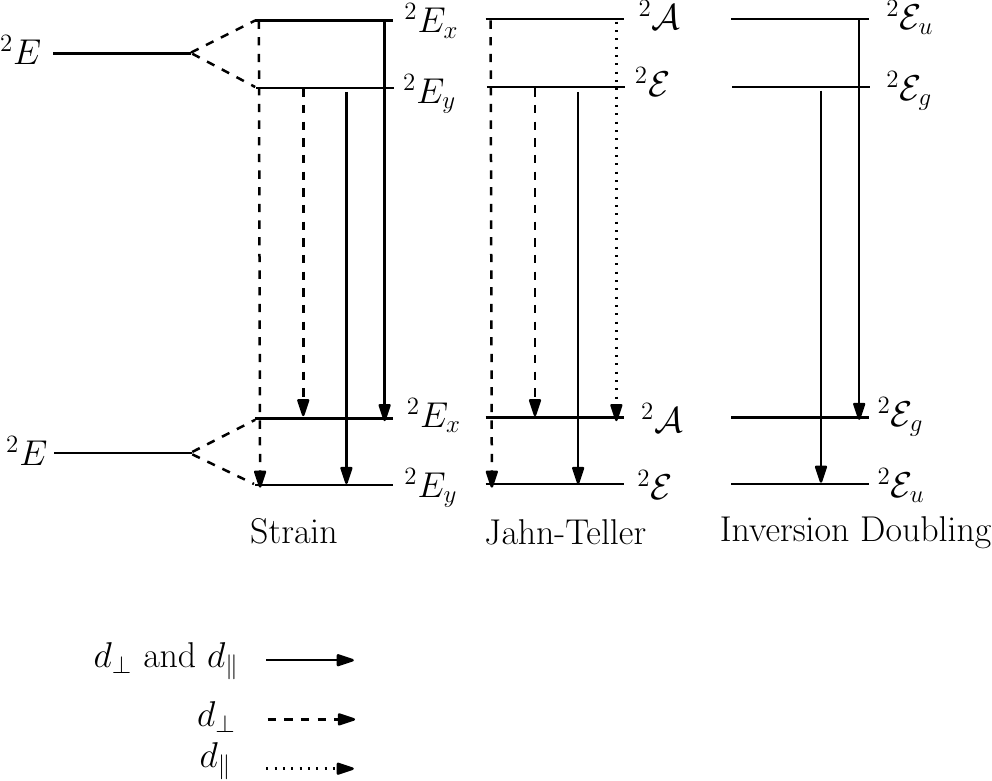}
\caption{
	The polarization selection rules of the alternate origins of the optical
	ZPL fine structure that are inconsistent with experiment: strain, the
	Jahn-Teller effect and inversion doubling. Strain lowers the symmetry of
	the center and splits both the orbitally degenerate ground and excited
	$^2E$ electronic levels into individual orbital levels denoted $E_x$ and
	$E_y$. The Jahn-Teller effect couples the $^2E$ electronic levels with $E$
	vibrational modes to yield lowest $^2{\cal E}$ and $^2{\cal A}$ vibronic
	levels \cite{stoneham2001theory}. Note that the relative intensities of the fine structure lines
	depends on the Jahn-Teller energy. In inversion doubling at a $D_{3d}$
	site, the $^2E_g$ and $^2E_u$ electronic levels are coupled with an $A_u$
	vibrational mode to yield odd $^2{\cal E}_u$ and even $^2{\cal E}_g$
	vibronic levels \cite{davies1976optical}.
}
\label{incorrect_fine_structure}
\end{figure}

Importantly, the observed selection rules are inconsistent with a strain origin
of the splittings (as illustrated in Figure \ref{incorrect_fine_structure}).
The negligible effect of strain is further supported by the observed
homogeneity of splittings in our sample. Figure \ref{incorrect_fine_structure}
also shows that our observed selection rules are inconsistent with the other
two previously proposed origins of the fine structure: the static
Jahn-Teller effect \cite{goss1996twelve-line}, and inversion doubling at a
site with $D_{3d}$ symmetry \cite{clark1995silicon}.

\section{Excitation polarization indicates a new electronic transition}

\begin{figure*}
\includegraphics[width=17.8cm]{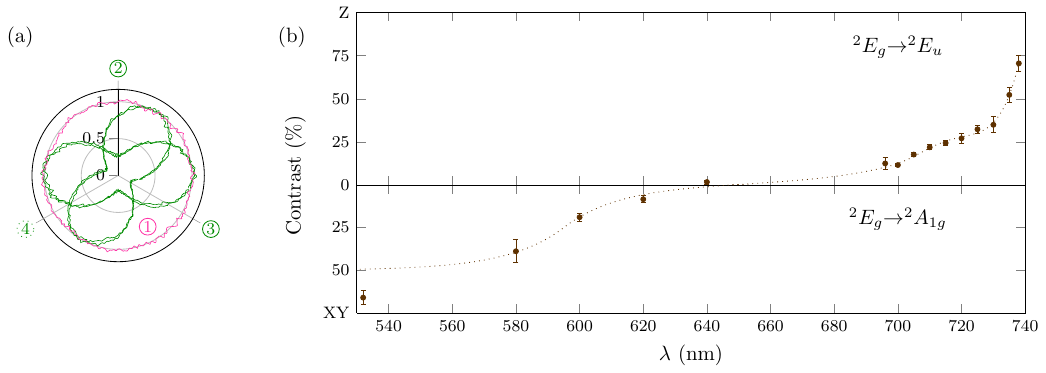}
\caption{
	(a) The PLE polarization dependence for a 532\,nm laser incident on the
	(111) surface.  These have been normalised to polarization along Y and the
	PLE contrast $C_{\text{PLE}}=66\pm4\%$ is in the opposite sense to that in
	PL.
	(b) The absorption contrast over a range of laser wavelengths. The error
	bars indicate the statistical variation of polarisation contrast measured
	over seven SiV$^-$ sites. The high contrast of the ZPL diminishes quickly
	for off-resonant excitation.  The contrast changes to the ``opposite''
	direction around 600\,nm, but the broad OPO line has smeared out this
	transition.   The dotted line is a visual guide only.
	}
\label{absorption}
\end{figure*}

Despite its weak emission phonon sideband that extends only $\sim$100\,nm
(200\,meV), PL of the 738\,nm ZPL is easily excited using 532\,nm light
(450\,meV above the extent of the absorption band).  In order to investigate
this unexpected phenomena, we measured the PLE polarization dependence at
532\,nm.  To perform these experiments the HWP and polarizer were moved from
the detection path to the incoming laser beam as illustrated in Figure
\ref{setup_drawing}.  Figure \ref{absorption}(a) shows the polarization contrast for
the three orientations seen in the $\{111\}$ face.  While the contrast seems
similar to the PL measurements in Figure \ref{zpl_polar_plots_111}, the oblique
orientations \circled{2} and \circled{3} have their PLE polarized opposite to
their PL.  Instead of having maxima along the X\,Z direction, the PLE
measurements show maxima in the Y direction.  Combined with the contrast of
about 70\%, this indicates that the excitation transition only involves
$d_\perp$ (as discussed in the previous section). 

For a $\langle111\rangle$ aligned site with $C_{3v}$ or $D_{3d}$ symmetry,
these dipole selection rules correspond to a transition between $E$ and $A$
electronic levels and imply the presence of a $^2\!A$ level.  Theoretical
calculations of the $D_{3d}$ split-vacancy model of the center predict such
excited $^2\!A_{1g}$ and $^2\!A_{2u}$ levels \cite{goss1996twelve-line,
dhaenens-johansson2011optical}. A $^2\!E\rightarrow{}^2\!A$ absorption from the
ground $^2E$ level that is accompanied by a significant phonon sideband will
explain the strong PLE observed at 532 nm. The observed PL at 738\,nm for
532\,nm excitation will occur if the $^2\!A$ level principally decays
non-radiatively to the excited $^2\!E$ level.

In order to estimate the energy of the $^2\!A$ level, the PLE polarization
contrast was measured in the range 532-640\,nm using a Ti:Sapph pumped optical
parametric oscillator (OPO) and in the range 696-725\,nm using a Ti:Sapph laser
(see Figure \ref{absorption}(b)). The PLE contrast switches from corresponding
to a $^2\!E\rightarrow{}^2\!E$ transition to a $^2\!E\rightarrow{}^2\!A$ transition
between 580 and 620\,nm. As the pulsed OPO produced wide-band excitation
($\sim$10\,nm), the precise wavelength where the contrast changes could not be
resolved, but it must occur around 600\,nm. This agrees well with a previously
reported PLE band assigned to SiV$^-$, which commences at 605\,nm (2.05\,eV)
\cite{iakoubovskii2000luminescence}. The PLE band was previously concluded to
be excitation into the diamond conduction band because of the absence of a ZPL
and broad similarities with the diamond ultraviolet absorption
band\cite{iakoubovskii2000luminescence}.  However, this was disputed by
subsequent \textit{ab initio} calculations, which place the electronic levels
of SiV$^-$ much deeper within the diamond bandgap \cite{goss1996twelve-line}.

Here, we assign that PLE band to an absorption transition from the ground
$^2\!E$ level to a $^2\!A$ level 2.05\,eV above.  We propose that the observed
absence of a ZPL implies that this is a  transition between electronic levels
of the same parity ($^2\!E_g$ and $^2\!A_{1g}$). Selection rules forbid purely
electronic transitions between levels of the same parity, but they do allow
transitions that involve the creation/annihilation of a phonon mode with odd
parity \cite{stoneham2001theory}.  This explains why a large PLE band can occur
with no ZPL, but it requires inversion symmetry.  This adds experimental
support for the conventional split-vacancy $D_{3d}$ model of SiV$^-$. 

It is known that other color centers in diamond have significant phonon
sidebands which accompany transitions between electronic configurations that
excite (quasi-)local phonon modes of $A$ symmetry \cite{davies1981jahn-teller}.
As the $D_{3d}$ split-vacancy structure permits odd parity $A_{2u}$ local
displacements of the silicon atom and its six nearest-neighbour carbon atoms,
it is very plausible that an $A_{2u}$ vibrationally-allowed
$^2\!E_g\rightarrow{}^2\!A_{1g}$ transition gives rise to the 605\,nm PLE band.
Given the small energy gap $\sim$0.37\,eV between the excited $^2E_u$ and
$^2\!A_{1g}$ levels, it is also plausible that rapid non-radiative
$^2\!A_{1g}\rightarrow{}^2\!E_u$ decay via a few phonons gives rise to the
$^2\!E_u\rightarrow{}^2\!E_g$ PL when the $^2\!E_g\rightarrow{}^2\!A_{1g}$
transition is excited \cite{rogers2010how}.


\section{Polarization of phonon sideband}

\begin{figure}
\includegraphics[width=8.6cm]{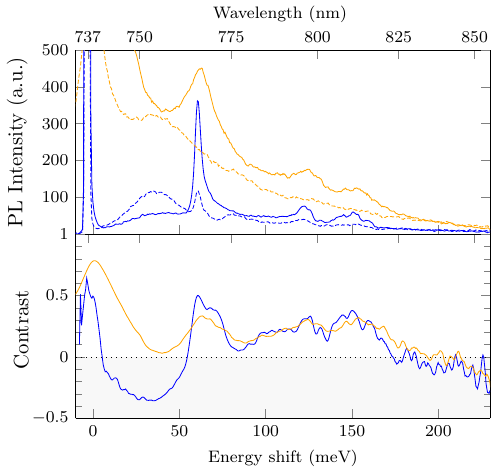}
\caption{
	Polarization of the PL sideband.
	(a) PL sideband at 8\,K (blue) and room temperature (orange) measured in
	the X\,Z direction (solid) and the Y direction (dashed).  There is
	considerable variation in polarization, with the 41\,meV peak being
	polarized oppositely to the ZPL.
	(b) The contrast $C$ for 8\,K (blue) and room temperature (orange).  The
	ZPL shows greater contrast at RT.
	}
\label{sideband_emission}
\end{figure}

Figure \ref{absorption}(b) demonstrates that the PLE polarization varies
considerably between excitation wavelengths 690\,nm and 737\,nm, which is
within the absorption sideband of the $^2\!E_g\rightarrow{}^2\!E_u$ transition.
The need for arbitrary laser wavelengths makes it challenging to examine this
absorption sideband in PLE.  Instead, the emission sideband was investigated
using PL measurements with 532\,nm excitation.  We measured the PL polarization
contrast within the emission phonon sideband at 8\,K and at room temperature (RT),
as shown in Figure \ref{sideband_emission}(a).  The features of this sideband
have been observed before \cite{clark19911.681, feng1993characteristics,
sternschulte19941.681-ev, iakoubovskii2000luminescence}, so the discussion here
will be restricted to their polarization. 

The first sideband peak (at 41\,meV) is entirely due to
$d_\perp$.  The sharp feature at 64\,meV is dominated by $d_\parallel$, like
the ZPL (Figure \ref{sideband_emission}(b)), which is consistent with it being
a local phonon mode of $A$ symmetry \cite{sternschulte19941.681-ev}.  The higher energy
phonon peaks also have similar polarization dependence to the ZPL, but with
lower contrast. At RT the phonon sideband peaks are less
distinct, but roughly the same polarization characteristics are observed.  The
41\,meV peak disappears as it merges with the shoulders of the ZPL and the
64\,meV peak, but its opposite polarization produces a small spectral region
with $C=0\%$.  Interestingly, the ZPL contrast increases to $C=80\%$ at RT, which
is of interest to RT applications using the center as a single photon source.

The variation of optical polarization within the phonon sidebands of the
$^2\!E_g\rightarrow{}^2\!E_u$ transition is strong evidence of dynamic
Jahn-Teller effects within the $^2\!E$ levels \cite{davies1976optical,fu2009observation}.

\section{Comparison with previous reports of SiV$^-$ alignment}

In this work, it has been established that the SiV$^-$ center is aligned along
the $\langle111\rangle$ crystal vectors.  This conclusion agrees with previous
\textit{ab initio} calculations  \cite{goss1996twelve-line} and related EPR
measurements of SiV$^0$ \cite{dhaenens-johansson2011optical}, and is further
confirmed by recent Zeeman measurements on low strain SiV$^-$ samples
\cite{hepp2014electronic}.  However, our result differs from previous optical
polarization \cite{brown1995site,neu2011fluorescence} and Zeeman
\cite{sternschulte1995uniaxial} studies.  In this section, we discuss these
previous studies in further detail.

Brown and Rand \cite{brown1995site} concluded a $\langle110\rangle$ alignment
with a single dipole along Z.  They measured SiV$^-$ ensembles in CVD films,
and the interpretation relied on assuming individual sites were randomly
oriented along equivalent crystal vectors.  We have seen evidence for
preferential alignment of SiV$^-$ sites incorporated during growth, and if
similar effects were present in their CVD films then it is difficult to
re-interpret the ensemble data.  A reinterpretation is also complicated since
they used 488\,nm excitation, which we have shown to probe the
$E_g\leftrightarrow A_{1g}$ transition.

Neu et al \cite{neu2011fluorescence} made measurements on diamond nanoislands
with $\{100\}$ surfaces, and deliberately avoided the 2.05\,eV excitation band.
They saw two main sets of orientations, aligned with the $\langle110\rangle$
nanoisland edges and separated by 90 degrees (their Figure 7).  When viewing
into a $\{100\}$ surface the $\langle111\rangle$ vectors are also aligned with
$\langle110\rangle$ crystal edges, and so their results are in fact consistent
with SiV$^-$ having a $\langle111\rangle$ orientation.  In addition, they saw
some SiV$^-$ sites angled at 45 degrees to the $\langle110\rangle$ nanoisland
edges (their Figure 9), and attributed them to $\langle110\rangle$ oriented
sites.  Still more sites were observed at seemingly arbitrary angles, however,
suggesting that it is difficult to identify the precise crystal alignment of
every nanoisland.

Their data indicate nearly 100\% contrast, which is significantly higher than
any measurement included here in Figures \ref{zpl_polar_plots_111} and
\ref{zpl_polar_plots_100}.  It is plausible that effects in the nanodiamond
environment alter the apparent dipole moments, and measured contrast is
dependent on background correction techniques.  Additionally, we have shown
that emission polarization varies significantly across the sideband, and does
reach high contrast on the ZPL at room temperature.  It is difficult to compare
previous reports without knowing the spectral band over which emission was
measured.

A $\langle100\rangle$ orientation was tentatively concluded from Zeeman
splitting results \cite{sternschulte1995uniaxial}.  Zeeman measurements probe
spin sublevels, which have not been addressed here.  More recent and thorough Zeeman
measurements have been subsequently reported that are compatible with a
$\langle111\rangle$ orientation \cite{hepp2014electronic}.

%
%

\section{Conclusion}

We have established the SiV$^-$ to be aligned with the $\langle111\rangle$
crystal vectors, having its strongest dipole moment along this symmetry axis.
Our results are consistent with the defect site having $D_{3d}$ symmetry and a
primary optical transition between two $E$ states.  Both of these $E$ states
are split at zero-field by spin-orbit coupling. An additional higher
$^2\!A_{1g}$ level allows efficient off-resonant excitation in a polarisation
orthogonal to the ZPL emission. At room temperature the ZPL polarisation
contrast increases from $C=60\%$ to $80\%$ and the ZPL approximates emission
from a single dipole.  These properties are of particular interest for
applications such as cavity QED and quantum entanglement algorithms that
require well polarized single photons.

\begin{acknowledgments}
The authors acknowledge funding from ARC (DP120102232), EU (DIAMANT), ERC,
German Science Foundation - DFG (SFB TR21, FOR1482, FOR1493), German Israeli
Foundation, JST, DARPA, Sino-German Center.
\end{acknowledgments}
L.R. and K.J. contributed equally to this work.
\bibliography{siv_electronic_structure}

\end{document}